\documentclass[prb,twocolumn,showpacs,floatfix,superscriptaddress]{revtex4}
\usepackage{eepic}
\usepackage{epsf}
\usepackage{amsmath}
\usepackage{graphicx}
\newcounter{saveeqn}%

\begin{document}

\narrowtext

{\bf Comment on ``Triplet-to-Singlet Exciton Formation 
in poly(p-phenylene-vinylene) Light-Emitting Diodes''}
\medskip

Lin {\it et al.} claim to have 
determined the ratio $\gamma$ of triplet to singlet 
excitons  in poly(p-phenylene-vinylene) (PPV) light 
emitting diodes by measuring the singlet and triplet 
absorption spectra under conditions of optical and 
electrical excitation \cite{Lin}. The authors find 
that (a) $\gamma >>$ 3 at weak electric fields, and 
(b) $\gamma < 3$ at moderate electric fields. This 
steep electric field dependence and $\gamma 
< 3$ at moderate fields is explained 
assuming :
(i) initial formation of a high energy triplet $T_2$ from
the triplet electron-hole (e-h) pair, which lies slightly 
below the e-h continuum and which is dipole allowed 
from the lowest triplet $T_1$, (ii) a phonon bottleneck 
\cite{Hong} in the nonradiative relaxation from $T_2$ to 
$T_1$, due to the large energy gap between them, and (iii) 
the competing field-induced dissociation of the $T_2$ exciton, 
and the consequent suppression of $\gamma$. The authors 
quote reference \onlinecite{Rohlfing} to state that since the 
higher energy singlet exciton $S_2$ is only 0.3 eV above 
$S_1$, phonon bottleneck and field induced dissociation are 
absent in the singlet channel.

The lowest excitations in PPV can be simulated within 
the Pariser-Parr-Pople model for linear polyenes with 
artificially large effective bond alternation \cite{Soos}, 
\begin{eqnarray}
\label{H_PPP} 
H = - \sum_{i, \sigma} t(1\pm\delta)
(c_{i \sigma}^\dagger c_{i+1,\sigma}+ h.c.) + \nonumber \\
U \sum_i n_{i \uparrow} n_{i \downarrow} +
\sum_{i<j} V_{ij} (n_i -1)(n_j -1)
\end{eqnarray}
where all terms have their usual meanings. Using the standard 
$U$ and t, Ohno parametrization for $V_{ij}$ and $\delta$ 
= 0.2 \cite{Soos} we have calculated the exact singlet and triplet
energies  
for the chain with N = 12 atoms. 
In Fig.~1, we show the exact 
triplet energy spectrum 
between $T_1$, hereafter 1$^3{\rm B}_u^-$, and the 
dipole-coupled $T_2$, hereafter 1$^3{\rm A}_g^+$. We have also 
included the optical singlet exciton $S_1$ (1$^1{\rm B}_u^-$) 
and 
the singlet exciton $m^1{\rm A}_g^+$,
which has the largest transition dipole with 
the $1^1{\rm B}_u^-$ within (1) (the quantum number $m$ is 
N-dependent and is 8 in N = 12) \cite{DGuo}, and is closer 
to the e-h continuum \cite{Shimoi} than the 1$^3{\rm A}_g^+$. 
The m$^1{\rm A}_g^+$ is observable in nonlinear 
spectroscopy \cite{DGuo,Liess} 
which places this state above
the $S_2$ state 
calculated in reference \onlinecite{Rohlfing} and cited by Lin 
{\it et al.} The $S_2$ state in reference \onlinecite{Rohlfing}
is the lowest two-photon state 2$^1{\rm A}_g^+$ of PPV and not 
the $m^1{\rm A}_g^+$.
The energy gap between the $m^1{\rm A}_g^+$ and the 
1$^1{\rm B}_u^-$ is at least as dense as that between the 
1$^3{\rm A}_g^+$ and the 1$^3{\rm B}_u^-$,
and hence there exist singlet excitons nearly degenerate with 
1$^3{\rm B}_u^-$. The large energy of the 1$^1{\rm B}_u^-$ and
the large energy gaps in Fig.~1
are finite size effects. 
With increasing N, absolute energies as well as energy differences 
decrease rapidly \cite{DGuo}
and additional levels 
appear
in the energy gaps.
Different 
theoretical treatments
have confirmed this picture 
of the long chain limit \cite{Abe,Race}.
\begin{figure}
\begin{center}
\centerline{\resizebox{2.4in}{!}{\includegraphics{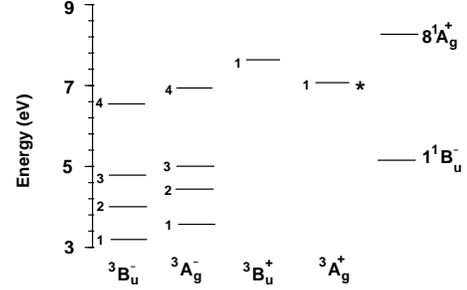}}}
\end{center}
\vspace{-0.5cm}
\caption[]
{The triplet energy spectrum between the 1$^3{\rm B}_u^-$ and the 
1$^3{\rm A}_g^+$ in a N = 12 chain
%%SR0404 adding
relative to the singlet ground state. 
%%end
Different symmetry subspaces are shown separately. 
%%SR0404 -- Modifying
%The asterisk indicates that the 1$^3{\rm A}_g^+$
State marked by asterisk 
%%end
is dipole-coupled to the 1$^3{\rm B}_u^-$. The 1$^1{\rm B}_u^-$ 
and the $m^1{\rm A}_g^+$ ($m$ = 8 in N = 12) are also included.}
\label{tnrg}
\end{figure}
\vskip 1pc
Two conclusions emerge from Fig.~1. First, the triplet 
spectrum between 
$T_1$ and the lowest dipole connected 
triplet, the so called $T_2$ in reference \onlinecite{Hong} 
is very dense in long chains. Absorption measurements miss all the
dipole--forbidden states in between.
Hence the phonon bottleneck in the nonradiative relaxation \cite{Lin,Hong} 
will not occur. 
The observed field dependence \cite{Lin} of $\gamma$ therefore 
cannot be explained within Lin et al.'s model. Second, the 
formation of the 1$^3{\rm B}_u^-$ triplet exciton via the 
$m^3{\rm A}_g^+$ would imply similar formation of the 
1$^1{\rm B}_u^-$ singlet exciton via the $m^1{\rm A}_g^+$. If 
indeed the energy spectra {\it were} sparse, contradicting
our calculations, then exciton dissociation would be larger 
in the singlet channel due to the closer proximity of the 
$m^1{\rm A}_g^+$ to the e-h continuum. This would have caused 
a field dependence of $\gamma$ opposite to what is observed by the 
Lin {\it et al.} \cite{Lin}. Our theoretical results therefore cast 
severe doubts on Lin {\it et al.'s} model, and by implication
on the experimental results themselves.

This work was partially supported by NSF and DST.\\
%\vskip 1pc

\noindent S. Mazumdar\\
Department of Physics \\
University of Arizona, Tucson, AZ 85721\\
%\vskip 1pc

\noindent Mousumi Das and S. Ramasesha\\
Solid State and Structural Chemistry Unit\\
Indian Institute of Science, 
Bangalore 560012, India.\\

\end{document}